\documentclass[conference]{IEEEtran}

\usepackage[cmex10]{amsmath}
\usepackage{amsfonts}
\usepackage{amssymb}
\usepackage{cuted,flushend}
\usepackage{multicol}
\usepackage[utf8]{inputenc}
\usepackage{bbm}
\usepackage{array}
\usepackage{graphicx}
\usepackage{float}
\usepackage{subfigure}
\usepackage{url}
\usepackage[table]{xcolor}

\newcommand{\pmin}{p^{\text{min}}}
\newcommand{\pmax}{p^{\text{max}}}
\newcommand{\fmin}{f^{\text{min}}}
\newcommand{\fmax}{f^{\text{max}}}
\newcommand{\argmin}{\text{argmin}}
\newcommand{\ac}{\mathcal{A}}
\newcommand{\acstar}{\mathcal{A}^*}
\newcommand{\achat}{\hat{\mathcal{A}}}
\renewcommand{\Re}{\mathbb{R}}
\newcommand{\rhostar}{\rho^*}
\newcommand{\rhohat}{\hat\rho}

\title{Learning for DC-OPF:\\Classifying active sets using neural nets}

\author{\IEEEauthorblockN{Deepjyoti Deka\IEEEauthorrefmark{1},
 and 
Sidhant Misra\IEEEauthorrefmark{1}}
\IEEEauthorblockA{\IEEEauthorrefmark{1}Theory Division, Los Alamos National Laboratory, NM, USA, 
$\{$deepjyoti, sidhant$\}$@lanl.gov}}

\begin{document}

\maketitle

\begin{abstract}
  The optimal power flow is an optimization problem used in power systems operational planning to maximize economic efficiency while satisfying demand and maintaining safety margins. Due to uncertainty and variability in renewable energy generation and demand, the optimal solution needs to be updated in response to observed uncertainty realizations or near real-time forecast updates. To address the challenge of computing such frequent real-time updates to the optimal solution, recent literature has proposed the use of machine learning to learn the mapping between the uncertainty realization and the optimal solution. Further, learning the active set of constraints at optimality, as opposed to directly learning the optimal solution, has been shown to significantly simplify the machine learning task, and the learnt model can be used to predict optimal solutions in real-time. In this paper, we propose the use of classification algorithms to learn the mapping between the uncertainty realization and the active set of constraints at optimality, thus further enhancing the computational efficiency of the real-time prediction. We employ neural net classifiers for this task and demonstrate the excellent performance of this approach on a number of systems in the IEEE PES PGLib-OPF benchmark library.
\end{abstract}
\section{Introduction}
The DC optimal power flow is a linear optimization problem routinely employed in power systems operation for finding the most economic generation dispatch and for market clearing. With the increasing integration of renewable energy resources such as wind and solar, as well as increasing complexity of demand-side behavior, there is a significant rise in the amount of uncertainty facing the system. In order to satisfy the demand while maintaining safe operational margins, system operators are employing larger and more frequent adjustments to the operating point in response to uncertainty realizations.

Traditionally, the necessary real-time adjustments to the generation is modelled using an affine control policy \cite{borkowska, vrakopoulou2012, roald2017corrective}, which mimics the behavior of the widely utilized automatic generation control (AGC). While affine policies are simple to handle computationally, they are restrictive, and can be sub-optimal in terms of cost and constraint enforcement \cite{roald2015optimal}. This motivates re-solving the OPF at a much faster time scale in response to uncertainty realizations and near real-time forecast updates. However, the tight latency requirements and large system sizes makes this a computationally challenging task. This has motivated recent literature that proposes the use of machine learning (ML) methods to learn the mapping from the uncertainty realization to the optimal solution of the OPF \cite{Ng2018-bj,misra2018learning}. This approach takes advantage of the fact that the OPF is routinely solved over and over again with varying values of renewable generation and demand, thus leading to readily available historical data for ML methods to train on. Once the ML model is trained, it can then be used to predict the optimal solution in real-time efficiently.

While it is possible to use off-the-shelf ML tools to learn the mapping from the uncertainty realization to the optimal solution, this approach suffers from a critical disadvantage - they struggle to produce predictions that satisfy the hard constraints in the OPF accurately. The main reason behind this drawback is that these ``black-box" methods do not exploit the fully known model of the power flows and structure of the DC-OPF problem. As a result a lot of training samples must be expended in order to learn the system model and satisfy constraints to an acceptable level of accuracy. To address this issue, \cite{Ng2018-bj,misra2018learning} adopted a new approach- instead of directly employing ML tools to learn the mapping, they propose learning the set of active constraints at optimality. The active set was shown to be a very effective intermediate learning step, which both simplifies the learning task as well as exploits known system and problem structure, and the approach performed very well for the DC-OPF problem, providing very high prediction accuracy for relatively low number of training samples for most systems.

A critical observation in \cite{Ng2018-bj,misra2018learning} is that, although the number of possible optimal active constraint sets is exponential in the size of the system, only a few of them are relevant for the OPF problem. Based on this observation, \cite{Ng2018-bj,misra2018learning} proposed an algorithm to identify these relevant active sets. Subsequently, for real-time prediction of the optimal solution, a simple exhaustive search strategy termed the ``ensemble policy" was proposed. The ensemble policy takes advantage of the fact that the DC-OPF is a linear program, and thus given the active set of constraints at optimality, obtaining the optimal solution needs a single linear operation corresponding to a single matrix-vector multiplication. Due to the exhaustive search involved, the computational complexity of the ensemble policy depends on the number of relevant active sets. When the number of active sets is low, the ensemble policy is an attractive choice owing to its simplicity. However its computational efficiency is compromised when the number of active sets is large since the exhaustive search requires one linear operation per candidate active set. 

In this paper we take a different approach in order to improve the computational efficiency of predicting the optimal active set. We employ classification algorithms from machine learning in order to learn the mapping from the uncertainty realization to the optimal active set. In the contexts of problems related to the optimal power flow, classification methods have been used in the literature to predict the behavior of locational marginal prices (LMPs) \cite{Geng2015-af}, and in \cite{Ardakani18} to predict whether a given constraint in an n-1 security constrained OPF belongs to the so-called class of umbrella constraints.

The goal in this paper is to use classifiers as an \emph{intermediate step} in learning the optimal solution to the OPF as a function of uncertainty realization. Directly attempting to learn this function requires estimation of a multi-dimensional continuous-valued function and is a challenging machine learning task. On the other hand, learning the map onto the active-set, which is a discrete and finite valued function is a task that is much simpler from the learning perspective, and one at which modern classification algorithms excel at.
In this paper we use classification methods based on neural networks \cite{GBC16} and demonstrate their performance through experiments on the IEEE PES PGLib-OPF v17.08 benchmark library \cite{pglib_opf}, by assessing the number of required training samples and prediction accuracy.

The main advantage in using classifiers over the ensemble policy proposed in \cite{Ng2018-bj} is that given a new uncertainty realization, the NN classifier can be directly used to predict the optimal active set. This prediction step is very fast, and bypasses the exhaustive search required in the ensemble policy. The performance of the classifier is given by the probability with which it predicts the correct optimal active set. The predictive performance can be further boosted by using the NN classifier to produce multi-rank or top-$K$ predictions. Given a new uncertainty realization, the neural net predicts a set of $K$ candidate active sets that has a high probability of containing the correct optimal active set while at the same time being much smaller than the set of all relevant active sets. Increasing the number of prediction classes boosts the prediction accuracy of the NN classifier, while at the same time significantly reducing the complexity of the subsequent exhaustive search based ensemble policy required to find the correct active set.

The rest of the paper is organized as follows. In Section~\ref{sec:formulation}, we describe the problem formulation and recap the active-set based approach to learning optimal solutions to the OPF. Section~\ref{sec:method} describes our classification method which we analyze using numerical experiments in Section~\ref{sec:experiements}. We conclude in Section~\ref{sec:conclusion} with directions for future work.
\section{Problem Formulation}  \label{sec:formulation}
In this section, we review the active-set approach to learning the optimal solutions introduced in \cite{Ng2018-bj} and then present the active set classification method. We keep same notation as used in \cite{Ng2018-bj} for consistency.

\subsection{The case for learning optimal power flow solutions} \label{subsec:why_learning}
We begin by stating the DC optimal power flow problem under uncertainty. 
\begin{subequations} \label{eq:opf}
\begin{align}
  \rho^*(\omega)\in\underset{p}{\text{argmin}}\ &c^\top p \label{eq:objective}\\
  \text{s.t.}\ &e^\top p = e^\top(d-\mu-\omega) \label{eq:balance}\\
	&\pmin\leq p\leq\pmax \label{eq:generation}\\
	&\fmin\leq M(Hp+\mu+\omega-d)\leq\fmax \label{eq:transmission}
\end{align}
\end{subequations}
Here $p$ denotes the vector of dispatchable generator set points, while $c$ is the vector of corresponding linear cost coefficients. $d$ is the vector of demands. $\mu$ represents the vector of forecast non-dispatchable active power (say from renewables), while $\omega$ denotes the uncertainty in the forecast. The vector $p^{min},p^{max}$ denote minimum and maximum generator limits. $e$ is the vector of all ones. $H$ is the mapping from generators to their respective buses, and $M$ denotes the matrix of power transfer distribution factors \cite{christie2000transmission}. Similarly, $f^{min}, f^{max}$ are the minimum and maximum transmission line flow limits. The objective \eqref{eq:objective} minimizes generation cost. Total Power balance, generation and transmission limits are enforced by \eqref{eq:balance}, \eqref{eq:generation}, and \eqref{eq:transmission}, respectively.

With growing uncertainty from renewable generation and fluctuating demand \eqref{eq:opf} needs to be solved at a much faster time scale in order to adjust generation in response to uncertainty realization. Traditionally, these real-time adjustments are modeled in the OPF using an affine policy \cite{borkowska, vrakopoulou2012, roald2017corrective}. However, the affine policy can be restrictive and is sub-optimal with respect to feasibility and optimality \cite{Ng2018-bj}. At the same time solving the OPF in real-time represents a significant online computational burden. To address this, \cite{Ng2018-bj,misra2018learning} proposed to use a machine learning approach to estimate the mapping from the uncertainty realization to the optimal solution (the function $\rho^*(\omega)$ in \eqref{eq:opf}) using offline computations. Given solution to the DC-OPF for multiple realization of the uncertainty in the form of $(\omega_1,\rho^*(\omega_1)),\ldots,(\omega_N,\rho^*(\omega_N))$, the goal of the learning task is to construct an estimate $\rhohat$ of the optimal mapping $\rhostar$. Note that such solutions are readily available using off-line optimization tools or from historical data given the frequency of OPF computations.

\subsection{Review of the active-set approach to learning solutions to the DC-OPF}
Directly constructing the estimate $\rhohat$ is a challenging task for off-the-shelf ML tools that fail to exploit the known structure and model description, and struggle to enforce constraints to acceptable levels of accuracy. Instead, \cite{Ng2018-bj,misra2018learning} proposed learning the set of active constraints at optimality as an intermediate step in the learning process. For a given realization $\omega$ the active set at optimality is defined as the set of constraints satisfied with equality by the optimal solution $\rhostar(\omega)$.
We follow the notation in \cite{Ng2018-bj} to represent the feasible set of \eqref{eq:opf} compactly in matrix notation as 
\begin{align}  \label{eq:feasibility_polytope}
  \mathcal{P}(\omega) = \{p: \ Ap \leq b+C\omega, \quad 
            e^\top p = e^\top(d-\mu-\omega)  \},
\end{align}
where
 \begin{align}
    A \!=\! &\left[ \begin{array}{r}
      \! I \\
      \! -I \\
      \! MH \\
      \! -MH
    \end{array}
    \right]\!\!\in\Re^{2(n+m)\times n}, \quad
    C \!=\! \left[ \begin{array}{r}
      \! 0 \\
      \! 0 \\
      \! -M \\
      \! M
    \end{array}
    \right]\!\!\in\Re^{2(n+m)\times v}, \nonumber \\[+2pt]
     b \!=\! &\left[ \begin{array}{l}
      \! ~~\pmax \\
      \! -\pmin \\
      \! ~~\fmax - M(\mu-d) \\
      \! -\fmax + M(\mu-d)
    \end{array}
    \right]\!\!\in\Re^{2(n+m)}. \nonumber
\end{align}
Here $n$ and $m$ denote the number of buses and lines in the transmission system considered. For an uncertainty realization $\omega$ the active set is given by $\ac(\omega) = \{i_1, \ldots, i_{n-1}\}$ corresponding to the indices of the rows of $A$ for which the inequality constraints are satisfied with equality. The active set approach in \cite{Ng2018-bj} for learning the optimal solution relies on two key observations: (i) Although the total number of possible active sets are exponential in the size of the system, for typical uncertainty distributions $\mathbb{P}_{\omega}$ of the random vector $\omega$, only a few of the active sets are realized (see Section \ref{sec:experiements} for numerical validation), and (ii) Since the optimal solutions to linear programs lie at extreme points of the feasible polytope in \eqref{eq:feasibility_polytope}, it is possible to obtain optimal solution from the knowledge of the active set using a single linear matrix operation. Let $B = \left[A_{\ac}^T e\right]^T$. Then the optimal solution is given by
\begin{align}  \label{eq:basis_policy}
  \rhostar(\omega) = \rho^{\ac}(\omega) = B^{-1} \left[ \begin{array}{c}
       b_{\ac} + C_{\ac} \omega \\
       e^\top(d-\mu-\omega) 
    \end{array} \right].
\end{align}

Based on the above observations, \cite{Ng2018-bj} proposes the following steps for learning the optimal solution to the OPF:
\paragraph{Identify important active sets} The set of relevant active sets for a given uncertainty distribution is identified by drawing sufficiently many samples from the uncertainty distribution $\mathbb{P}_{\omega}$ to obtain samples $\omega_1, \ldots, \omega_N$ and observing the corresponding optimal active sets $\ac(\omega_1), \ldots, \ac(\omega_N)$. In \cite{misra2018learning}, a stopping criterion is described to ensure that the number of samples $N$ is sufficiently large to discover most of the important active sets denoted by $I$.
\paragraph{Prediction using exhaustive search} Once the set of important active sets $I$ has been determined, for a new uncertainty realization $\omega$, the optimal solution can be obtained by the so-called \emph{ensemble policy} given by
\begin{align} \label{eq:ensemble_policy}
    \rhohat(\omega) :=\underset{\rho^{\ac}(\omega): \ac\in I}{\argmin}\ &c^\top \rho^{\ac}(\omega) \\
  	\text{s.t.}\ &\rho^{\ac}(\omega)\in \mathcal{P}(\omega) \nonumber
\end{align}
Essentially, the ensemble policy performs the linear operation in \eqref{eq:basis_policy} for each candidate active set $\ac$ in the set $I$ of important active sets obtained in step $(a)$, and the chooses the one which has the least cost and is feasible. As long as the set $I$ contains the optimal active set, the estimate $\rhohat$ provided by the ensemble policy is optimal and equal to $\rhostar$.

\subsection{Learning optimal solutions by classification of active sets}
The computational complexity of the ensemble policy is dictated by the number of important active sets $|I|$. While the simplicity of the ensemble policy makes it an attractive choice for systems for which $|I|$ is small, the computational burden of the prediction step can be large when $|I|$ is large, defeating the main motivation behind constructing the estimator $\rhohat$ for optimal solutions. 

In this paper, we employ classification algorithms from machine learning to directly learn the mapping from the uncertainty realization $\omega$ to the corresponding optimal active set $\ac^*(\omega)$.
For this purpose, we use a neural net classifier, the details of which are described in Section~\ref{sec:method}. The trained classifier can be used for both single-rank and multi-rank prediction.
\paragraph{Single rank prediction} The classifier is used to learn and predict the optimal active set for a given uncertainty realization. Specifically, it constructs an estimate $\achat(\omega)$ of the true function $\acstar(\omega)$. The performance of the classifier is assessed by its prediction accuracy given by
\begin{align}  \label{eq:singleclass_prediction_accuracy}
  \eta = \mathbb{P}_{\omega}(\achat(\omega) = \acstar(\omega)).
\end{align}
The prediction step of the classifier, which corresponds to evaluating the function $\achat(\omega)$ for a new uncertainty realization $\omega$ is highly computationally efficient. The estimate for the optimal solution $\rhohat(\omega)$ is then obtained by the linear operation in \eqref{eq:basis_policy} using the predicted active set $\achat(\omega)$:
\begin{align}  \label{eq:singleclass_rhohat}
  \rhohat(\omega) = \rho^{\achat(\omega)}(\omega).
\end{align}
\paragraph{Multi Rank/ Top-$K$ prediction} In this case, the classifier constructs an estimate consisting of a set of $K$ candidate active sets $\hat{I}_K(\omega) = \{\achat_1, \ldots, \achat_K\}$. The estimate for the optimal solution is then obtained by using the ensemble policy over the reduced set $\hat{I}_K$ of candidate active sets:
\begin{align}
  \rhohat(\omega) = \rho^{\hat{I}_K(\omega)}(\omega).
\end{align}
The rationale behind using a multi-class prediction is that increasing the number of predicted active sets from $1$ to $K$ boosts prediction accuracy, i.e., $\mathbb{P}_{\omega}(\acstar \in \hat{I}) > \mathbb{P}_{\omega}(\acstar = \achat)$. At the same time $K$ is chosen to be much smaller than $|I|$, and hence the complexity of the ensemble policy is reduced. The parameter $K$ can be chosen to obtain the desired level of trade-off between prediction accuracy and prediction complexity, both of which increase with increase in $K$.

\begin{figure}[tb]
\centering
\includegraphics[width=0.45\textwidth]{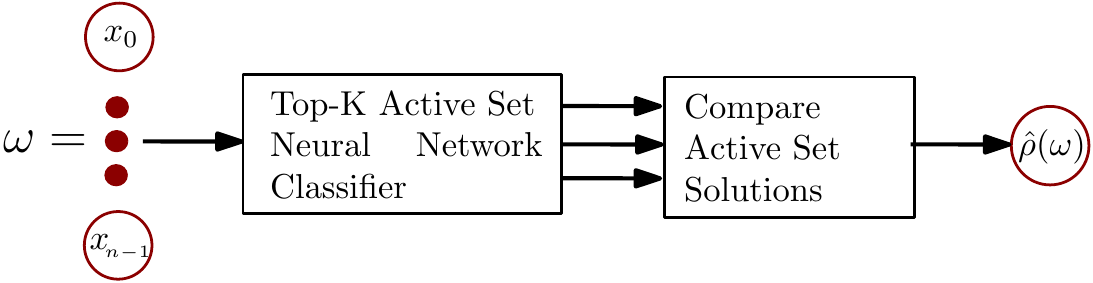}
\caption{OPF Solution via Active Sets}
  \label{fig:method}
\end{figure}

The overall solution strategy proposed is depicted in Fig.~\ref{fig:method}.


\section{Classification}\label{sec:method}
As described previously, the number of OPF active sets observed for each test system is only a small fraction of the total number that can be constructed. To learn the mapping from the vector $\omega$ of uncertainty to the set of active sets, we train a non-linear neural network based classifier. Note that as DC-OPF is a LP, a neural network may be an unnecessary sophistication for the classification task. However we consider this approach as we plan to extend it to harder non-linear variations of the problem, including AC OPF, OPF with chance and integer constraints etc. This work would thus serve as a benchmark and starting point for multiple directions of future work. It is also noteworthy that current neural network architectures are user-friendly for code-design and hence amenable for easy testing and implementation.
\begin{figure}[tb]
\centering
\includegraphics[width=0.45\textwidth]{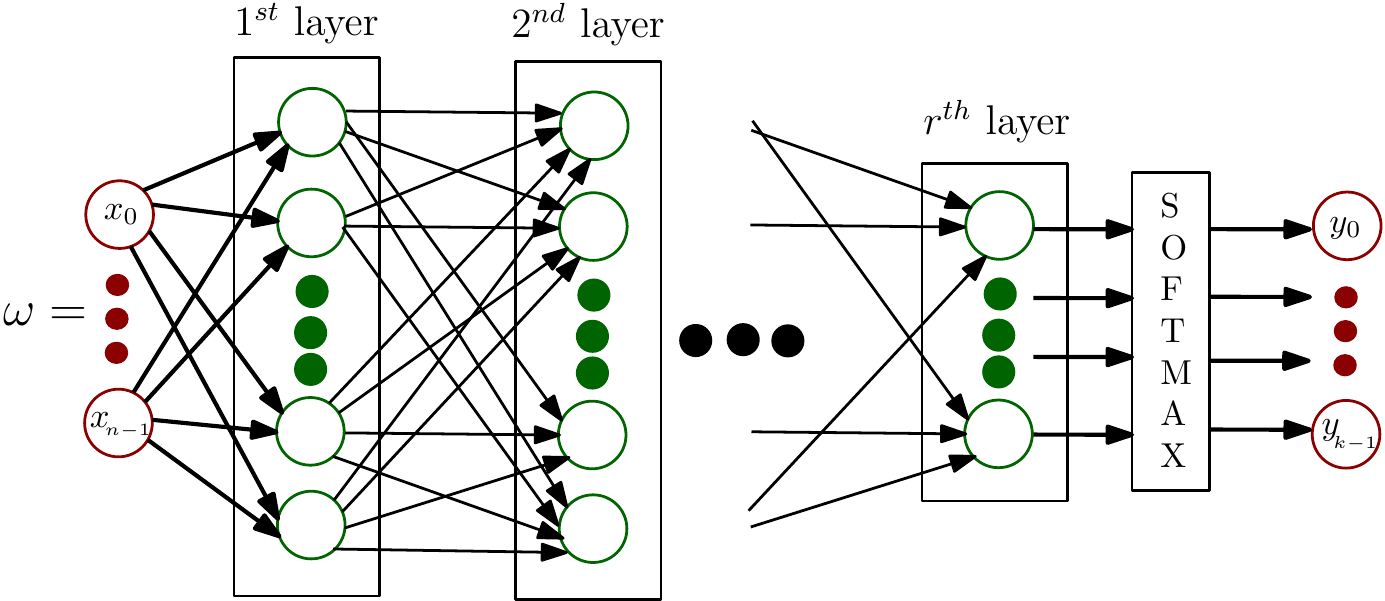}
\caption{Architecture of the Neural Network classifier}
  \label{fig:NN}
\end{figure}

\subsection{NN classifier}
We use a simple multi-layer fully connected neural network (NN) \cite{GBC16} for our classification task with nodal injections as input features. The generic structure of the NN is shown in Fig.~\ref{fig:NN}. The feature vector of nodal injections $\omega$ serve as the input to the first layer $\vec{x}^0 = [x_0, \dots, x_{n-1}]^T$. Each layer, except the final layer, in the NN comprises of a set of ReLUs (Rectified Linear Units) which takes as input, the output $x^j$ of the previous layer and conducts the following operation.
\begin{align}\label{out}
\vec{x}^{j+1} & = g(W_j^T \vec{x}^j + B_j)
\end{align}
where $W_i, B_i$ are respectively the output kernel and the bias of the $i^{th}$ layer. $g(\cdot) $ is the non-linear ReLU function
\begin{align}\label{relu}
g(x^j_k) &= \max(x^j_k, 0)
\end{align}
The final layer coverts its input $x$ to probability scores $\vec{y}$ used in classification using the following softmax function 
\begin{align}\label{softmax}
    y_i = \frac{e^{x_i}}{\sum^{k-1}_{i=0} e^{x_i}}
\end{align} for $i=0, \cdots, k-1$ where $k$ is the number of candidate active sets. 

\subsection{Training Process} 
We consider a training set $\mathcal{S}$ of size $N$ with known feature vectors $\omega^j$ and corresponding active set indices $\acstar(\omega^j)$. Let the set of all parameters in the NN be $\Theta$. We determine the optimal $\Theta$ is found by minimizing the following cross-entropy loss function \cite{GBC16} over the training set 
\begin{align}\label{loss}
l(\Theta ) & =  \frac{1}{N} \Sigma_{j=1}^N \Sigma_{i=0}^{k-1} {y}^{j }_i \log{f_{\Theta}( {\omega}^j)}
\end{align} 
where $f$ is the NN map from $\omega$ to the vector of output probabilities. $y^{j}$ is a $k$ length binary vector that takes a value of $1$ for the true active set $\acstar(\omega^j)$ and zero otherwise. The cost function thus represents a divergence cost between the true and estimated probabilities of the active set labels. To solve this optimization problem, we use a variant of the stochastic gradient descent method called Adam \cite{KBJ14}. 

In addition, for each ReLU layer in the NN, we use ``batch normalization'' \cite{ISC15} to eliminate the issue of covariance shift and include ``Dropout" \cite{srivastava2014dropout} during training for regularizing the parameters to improve the output. Note that the number of ReLU units and number of layers are hyper parameters that can be changed before executing the NN learning process, but not optimized inside it. After the training stage, we use the NN classifier on test data where the classification output for given input feature-set comprises of the top $K$ indices with the highest estimated probabilities. For $K=1$, it reduces to the standard multi-class classification output. In the next section, we discuss numerical results for active set classification for different test cases along with effects on classifier performance due to NN hyper parameters.

\section{Experiments}  \label{sec:experiements}
For our experiments, we consider $9$ different test cases from the IEEE PES PGLib-OPF v17.08 benchmark library \cite{pglib_opf}. The test cases are listed in Table~\ref{tab:case} along with their details. Note that the number of generator constraints is greater than the number of buses due to presence of co-located generation. Following the setting in \cite{Ng2018-bj}, we assume that the loads are uncertain, and that $\omega$ follow a multivariate normal distribution with mean zero and standard deviation $\sigma$ \footnote{Note that the choice of distribution is arbitrary and our method can be applied to any other distribution since it only relies on samples} proportional to the load, i.e., $\sigma_i = 0.03*d_i$.
The last column shows the total number of active sets observed by solving the DC-OPF for 50,000 independent samples drawn from the above distribution.
Only four test cases with $24, 57, 162,300$ buses (highlighted in Table~\ref{tab:case}) have greater than three active sets. As we include top-$3$ classification in our experiments, we restrict ourselves to these four test cases as results for the others will trivially be $100\%$ accurate. 

\begin{table}[ht]
\renewcommand{\arraystretch}{1.3}
\caption{Test-cases for learning}
\label{tab:case}
\centering
\begin{tabular}{|c|c|c|c|}
\hline
Test-Case & Generator Constraints & Flow Constraints & Active Sets \\ \hline
\textbf{24-bus} & 57&77 & \textbf{7} \\ \hline
30-bus  & 36 & 83 & 1\\ \hline
39-bus  & 49 & 93 &2 \\ \hline
57-bus & 64 & 161 & 3 \\ \hline
\textbf{73-bus}  & 172 & 241 & \textbf{24} \\ \hline
118-bus  & 172& 373 & 3 \\ \hline
\textbf{162-bus} & 174 & 569 & \textbf{17} \\ \hline
\textbf{300-bus}  & 369& 823 & \textbf{78} \\ \hline
1888-bus  & 2178 & 5063 & 3 \\ \hline
\end{tabular}
\end{table}
\begin{figure*}[tbh]
\centering
\subfigure[$24$-bus]{\label{24bus_sets}\includegraphics[width=0.248\textwidth]{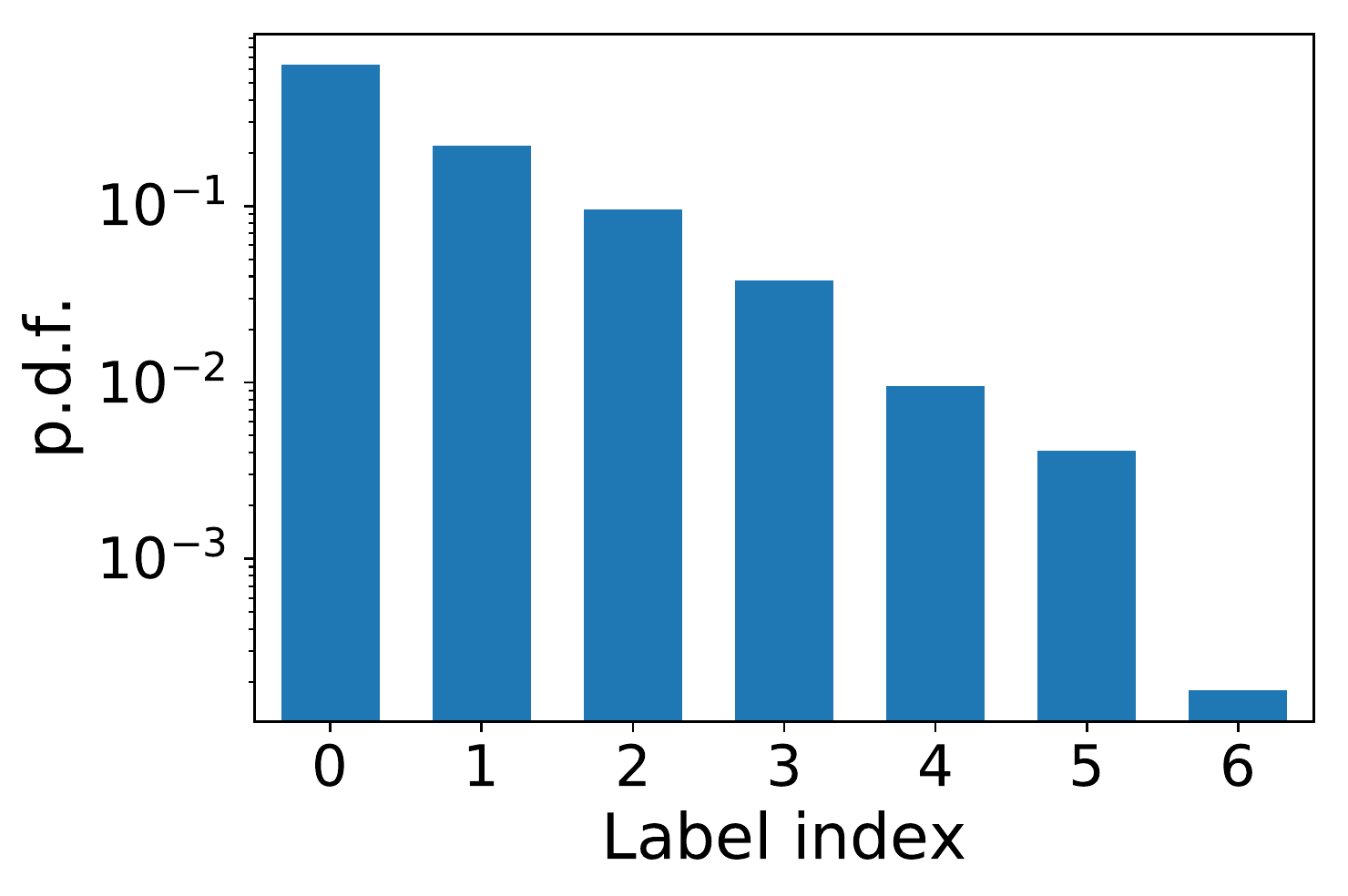}}\hfill
\subfigure[$73$-bus]{\label{73bus_sets}\includegraphics[width=0.248\textwidth]{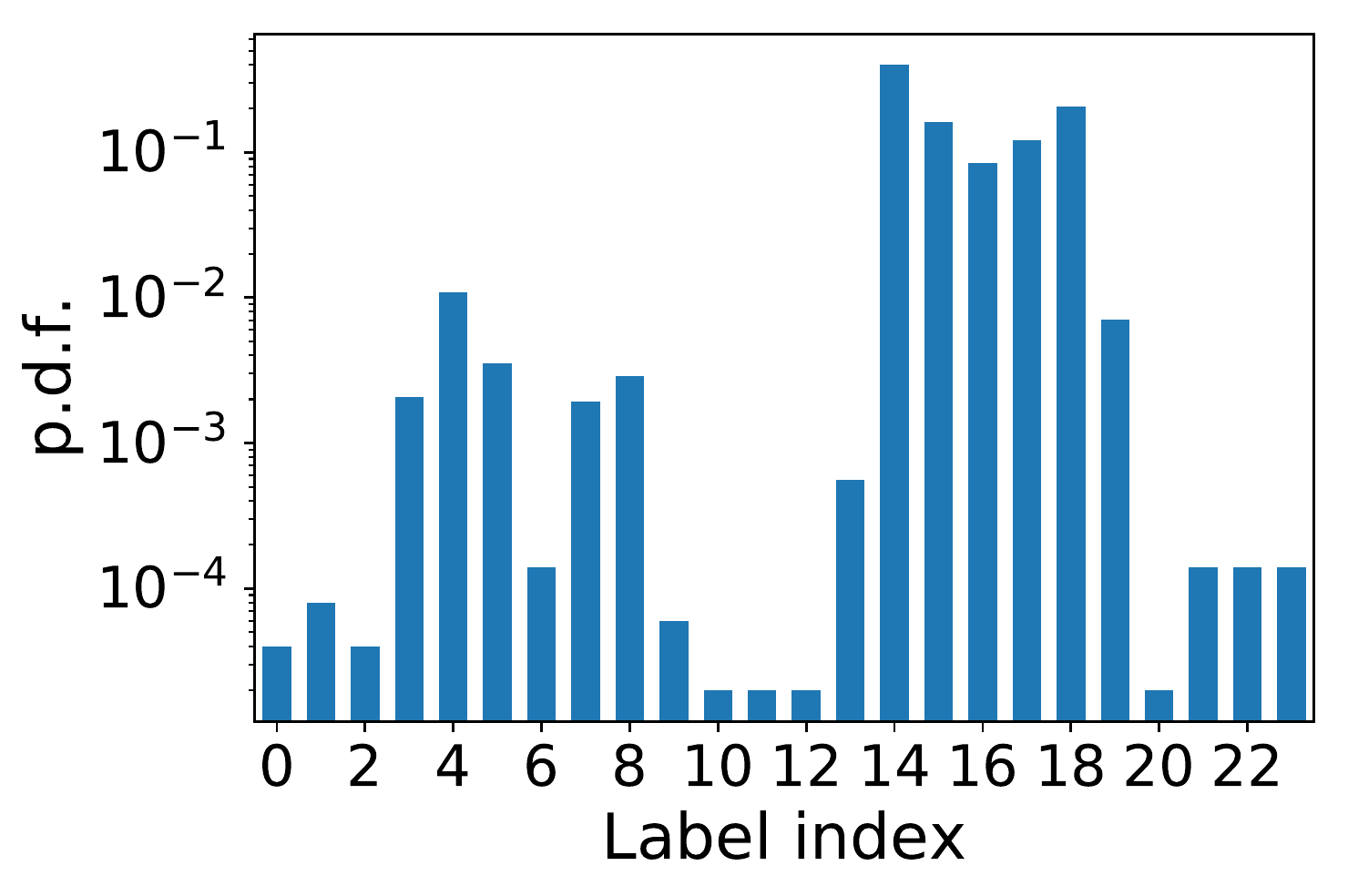}} \hfill 
\subfigure[$162$-bus]{\label{162bus_sets}\includegraphics[width=0.248\textwidth]{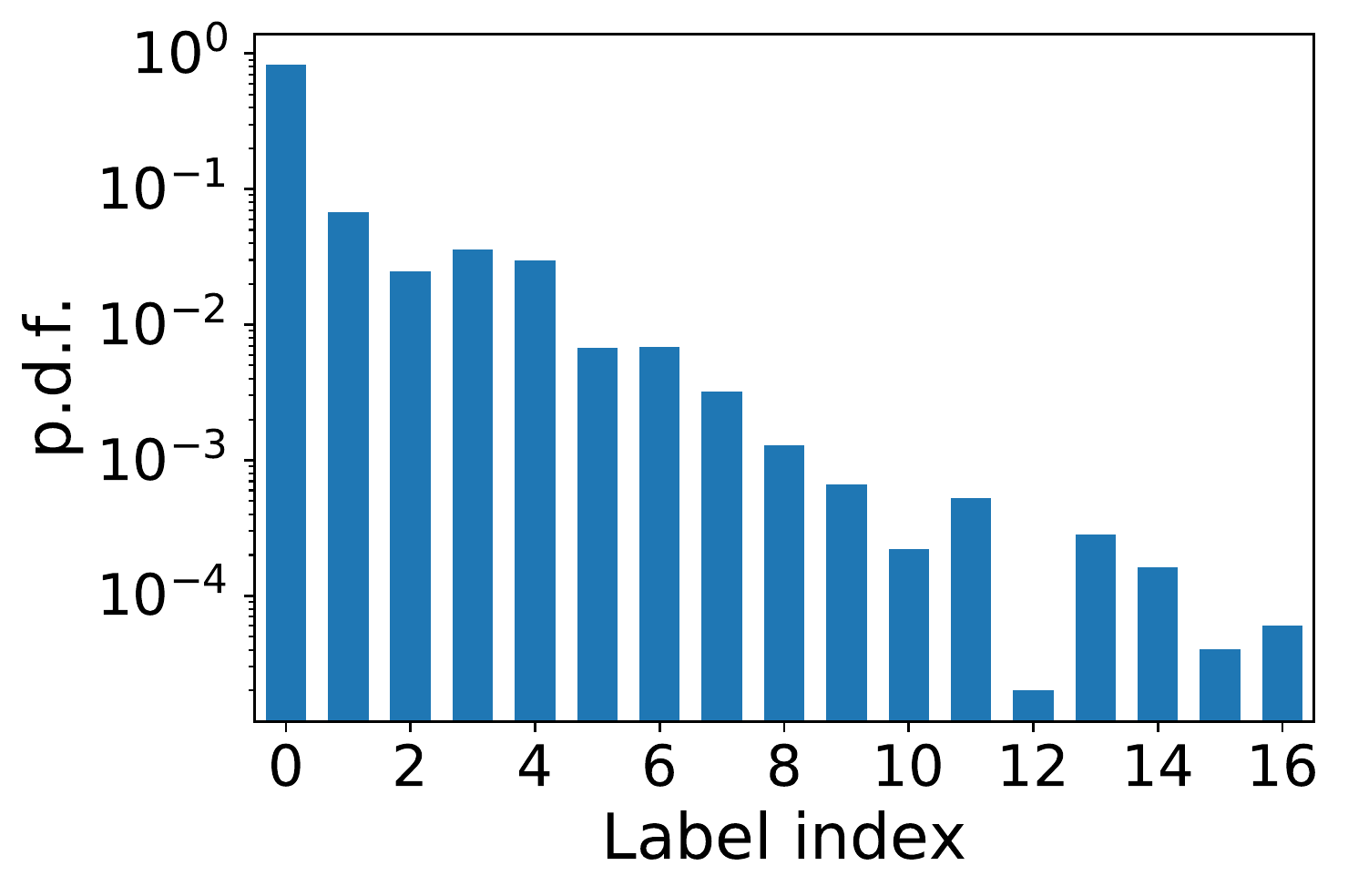}}\hfill
\subfigure[$300$-bus]{\label{300bus_sets}\includegraphics[width=0.248\textwidth]{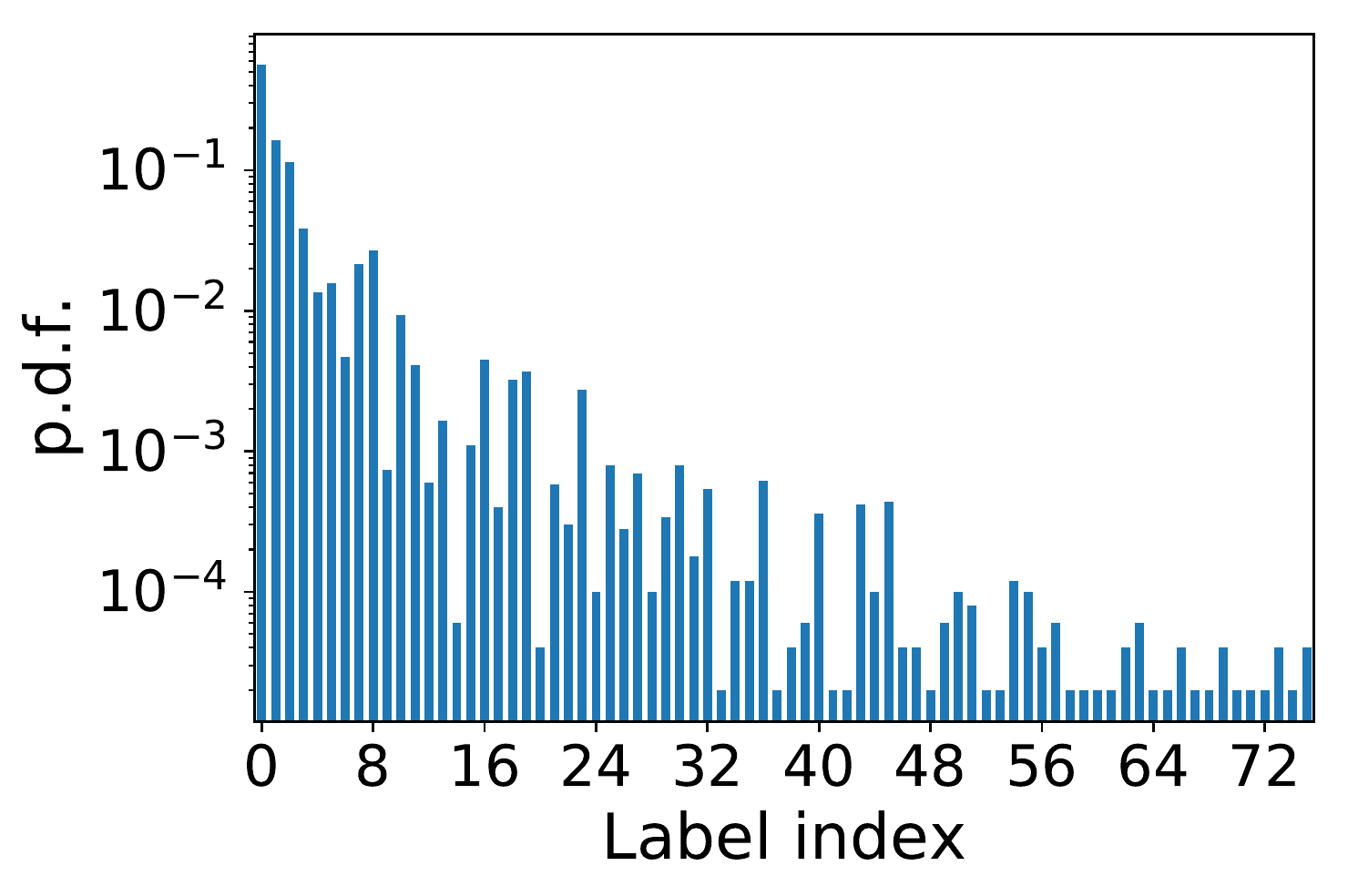}}
\caption{Probability distribution of active sets of different OPF test-cases}
  \label{fig:active_sets_pdf}
\end{figure*}

The percentage of active sets being significantly less than the maximum number possible can be partially explained by the fact that majority of the generator and flow constraints in the OPF samples for each test-system have a fixed status (either unconstrained, or constrained at upper or lower limit). This can be validated from  Table~\ref{tab:status}. In addition, variations in constraint statuses are correlated leading to further decrease in the total number of active sets.
\begin{table}[ht]
\renewcommand{\arraystretch}{1.3}
\caption{Percentage of constraints with fixed status in data-set}
\label{tab:status}
\centering
\begin{tabular}{|c|c|c|}
\hline
Test-Case & Fixed generator constraints & Fixed flow constraints\\ \hline
24-bus & 87.72\% &100\% \\ \hline
73-bus  & 96.5\% & 100\% \\ \hline
162-bus & 97.7\% & 98.6\% \\ \hline
300-bus  & 97.3\% & 99.2\% \\ \hline
\end{tabular}
\end{table}

For the selected test cases, the frequency or distribution of different active sets is shown in Fig.~\ref{fig:active_sets_pdf} over a sample space of $5\times 10^4$ samples each. Note that there are no immediate trends that can be observed in the distribution. While active sets with extremely low frequency (in single digits) exist, there are also multiple active sets that appear with comparable high frequency. These high frequency sets play a greater role in the empirical accuracy of our learning algorithm. We now discuss the performance of our NN architecture in classifying the correct active set. We use Tensorflow with Keras python \cite{chollet2015keras} package for our NN experiments. The NN parameters are optimized over the cross-entropy cost in Eq.~(\ref{loss}) on training data for $20$ iterations with a batch size of $32$ samples. This takes approximately 30 seconds for each test-case considered. Computing the scores over the test data involves a sequence of direct multiplications and hence takes less than $1$ sec in each case. We quantify the accuracy as the average fraction of test samples where the the true active set is among the top $K$ estimated ones. 

For the first set of experiments, we consider each test-case and look at the performance of single and multi-ranked classification ($K\leq3$) with different depths in the NN. For the deepest NN, we consider a $5$ fully-connected layers with $256,256,128,128,64$ ReLU units respectively. We use $4 \times 10^4$ samples for training and $10^4$ samples for testing for each highlighted test-case in Table~\ref{tab:case}. The results are presented in Table~\ref{tab:layer}. Notice that the performance does not degrade with the depth of the NN. This is not surprising given that for DC-OPF, the regions under each active set can be linearly separated. Hence few ReLU hidden layers may be sufficient. Further top-$3$ classifier for several cases gives close to $100\%$ accuracy, demonstrating that even if the label with highest estimated probability may be inaccurate, the true active set label is gets a sufficiently high score. The reduced performance for top-$1$ classifier for the $73$ bus test case compared to the other cases and classifiers can be related to the fact that multiple active set labels have equally high frequency (see Fig.~\ref{fig:active_sets_pdf}). This in turn is often the result of degeneracy in the optimal solution induced by co-located generators with identical costs. In such cases, there exist multiple generator set-points and active sets with identical OPF total cost. In the future, we will analyze the effect of removing degeneracy in generator set-points and its effect on active set classification.

\begin{table}[ht]
\renewcommand{\arraystretch}{1.3}
\caption{Accuracy of NN with different fully connected layers}
\label{tab:layer}
\centering
\begin{tabular}{|c|c|c|c|c|}
\hline
No.of layers: & 2 & 3 &4 & 5\\
\hline
\multicolumn{5}{|c|}{24-bus system} \\
\hline 
Top-$1$ &98.5\% &98.5\% &98.2\% &98.7\%\\ \hline
Top-$2$ &100\% &100\% & 100\% &100\%\\ \hline
Top-$3$ &100\% &100\% &100\% &100\%\\ \hline
\multicolumn{5}{|c|}{73-bus system} \\
\hline 
Top-$1$ &63.2\% &62.5\% &63.1\% &63.6\% \\ \hline
Top-$2$ &96.6\% &96.3\% &96.3\% &96.7\% \\ \hline
Top-$3$ & 98.8\% &98.7\% &98.8\% &98.9\%\\ \hline
\multicolumn{5}{|c|}{162-bus system} \\
\hline
Top-$1$ &96.2\% &96.6\% &96.5\% &96.9\% \\ \hline
Top-$2$ &99.7\% &99.6\% &99.6\% &99.6\%\\ \hline
Top-$3$ &99.9\% &99.9\% &99.9\% &99.9\%\\ \hline
\multicolumn{5}{|c|}{300-bus system} \\
\hline 
Top-$1$ &94.7\% &94.4\% &93.6\% &93.7\%\\ \hline
Top-$2$ &99.3\% &99.3\% &98.9\% &99.1\%\\ \hline
Top-$3$ &99.6\% &99.7\% &99.5\% &99.6\%\\ \hline
\end{tabular}
\end{table}
\begin{figure*}[!tb]
\centering
\subfigure[$24$-bus]{\label{24bus_samples}\includegraphics[width=0.248\textwidth]{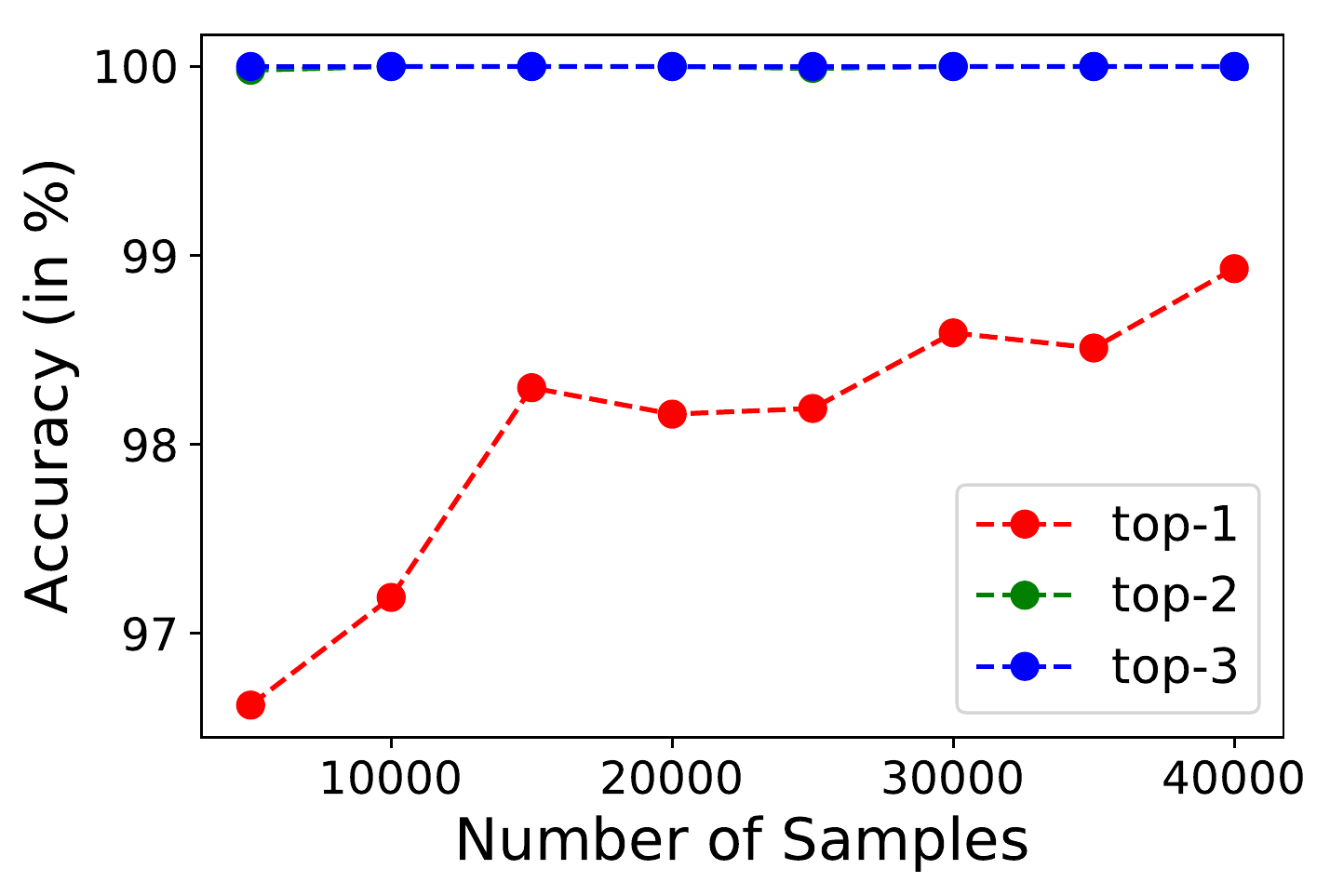}}\hfill
\subfigure[$73$-bus]{\label{73bus_samples}\includegraphics[width=0.248\textwidth]{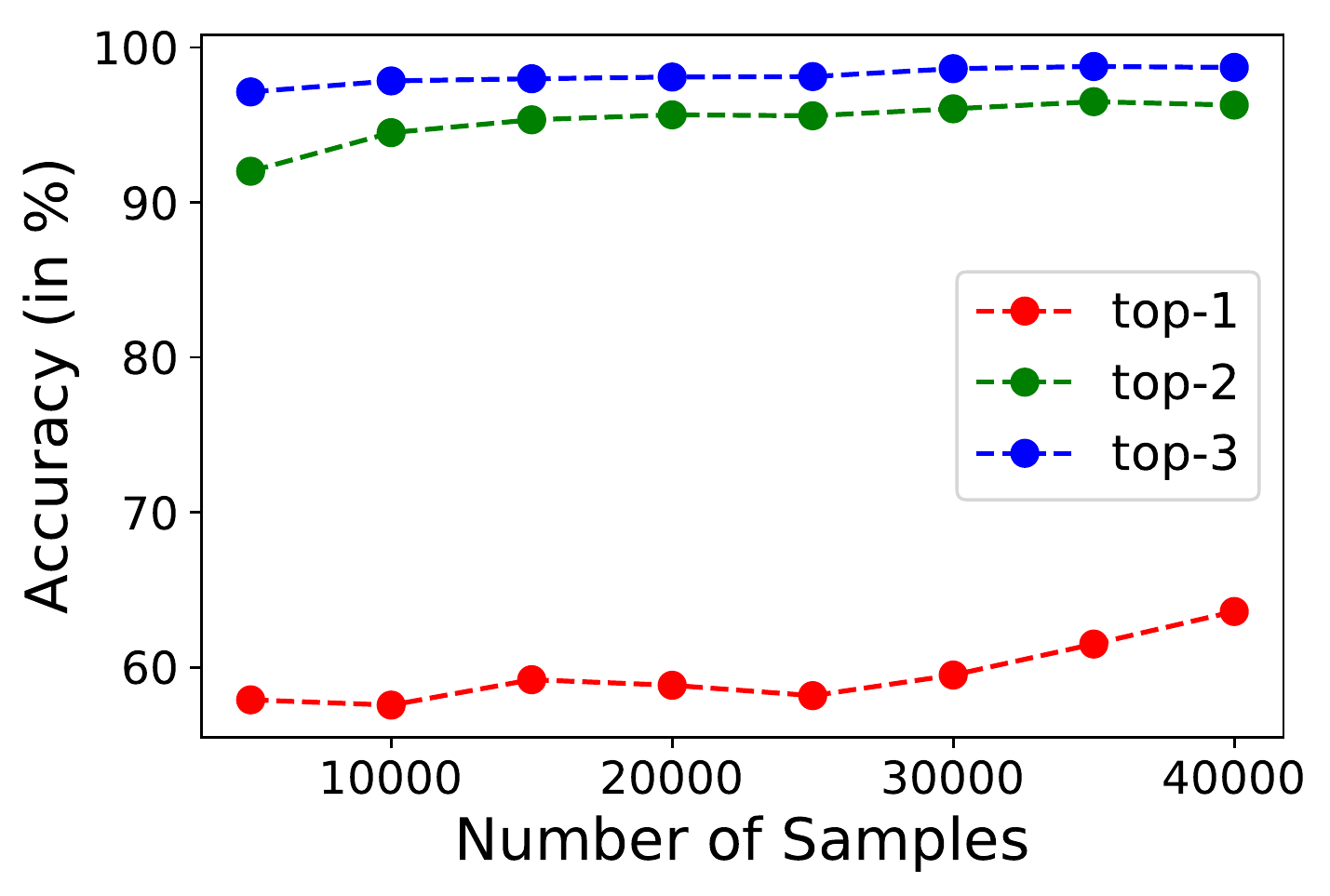}} \hfill
\subfigure[$162$-bus]{\label{162bus_samples}\includegraphics[width=0.248\textwidth]{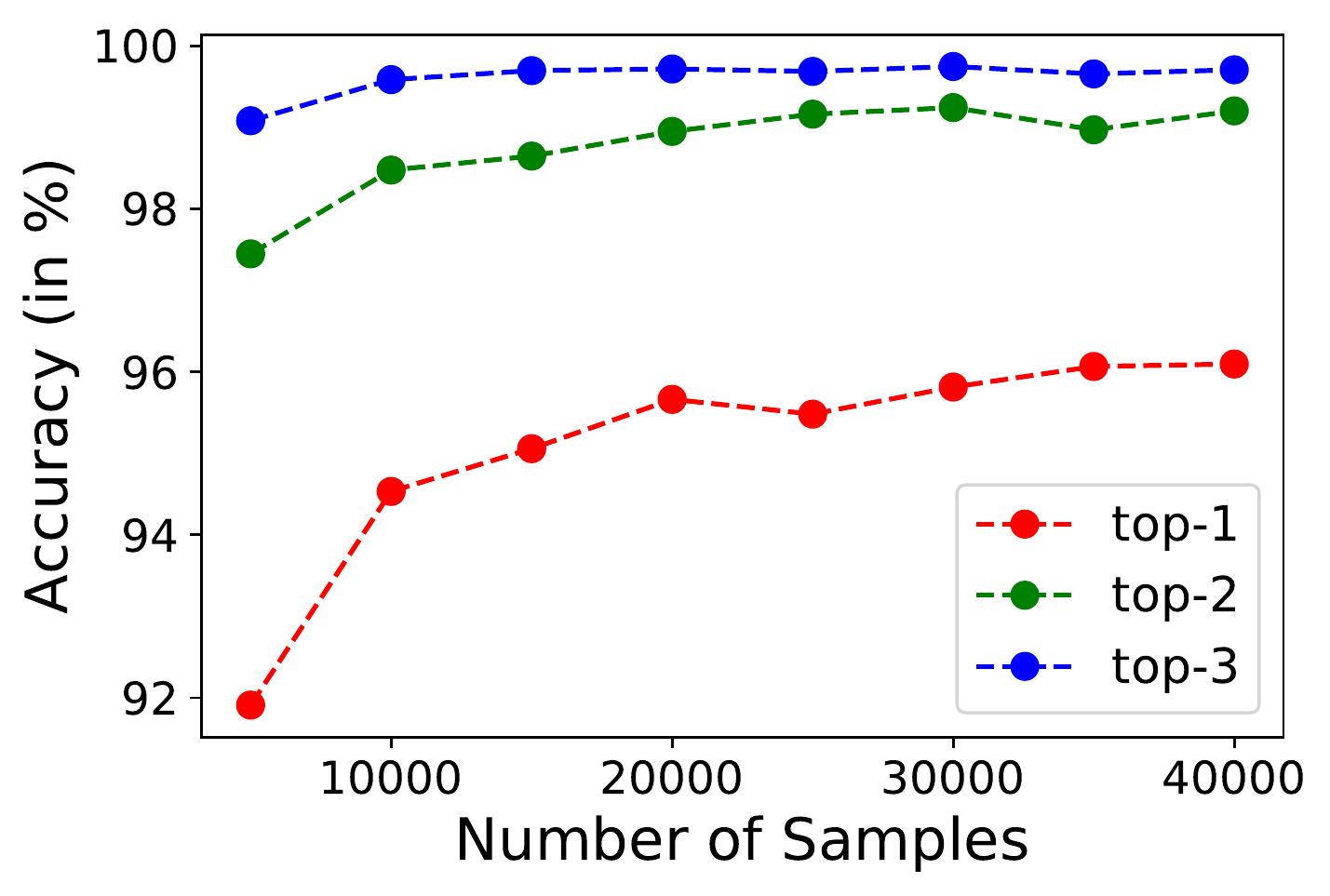}}\hfill
\subfigure[$300$-bus]{\label{300bus_samples}\includegraphics[width=0.248\textwidth]{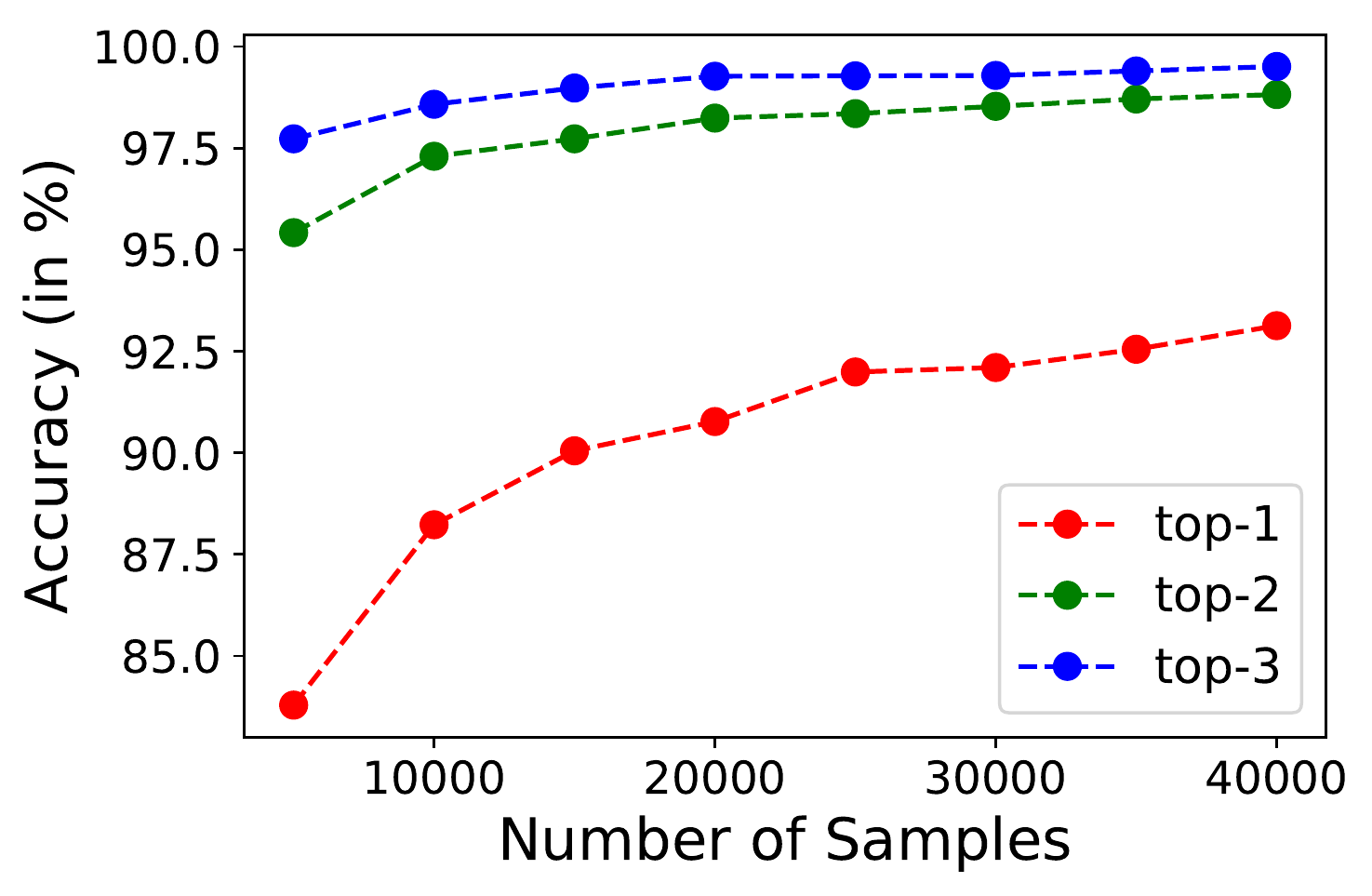}}
\caption{Accuracy of active set top-$K$ classification for different test cases with change in training data set size. The test data set is taken as $10000$ samples.}
  \label{fig:active_sets_sample}
\end{figure*}
Next, we experiment with the number of samples available for training a NN with $5$ ReLU hidden layers and study its effect on classification accuracy. For each of the four highlighted test cases in Table~\ref{tab:case}, we fix the number of test samples at $10^4$ and change the number of training samples from $5\times 10^3$ to $4\times 10^4$ samples. We report classification accuracy for top $1,2$ and $3$ ranked outputs in Fig.~\ref{fig:active_sets_sample}. Notice that the performance increases with $K$ and reaches almost $100\%$ for $K=3$ for all test cases. There is also a significant improvement in performance for all cases and sample sizes for $K = 2,3$ over $K=1$. This justifies our decision to consider multiple ranks for the classification task instead of using only the top-rank. In future work, we plan to investigate theoretically the sharp improvement in performance from $K = 1$ to $K = 2$. It can also be observed that for majority of cases, the improvement with sample sizes is higher for $K = 1$ than for $K=2,3$. Finally it is noteworthy that for the $73$-bus test system, the accuracy for top-$1$ classification is much lower (~$60\%$) than that for other test-cases (~$90\%$). However for $K = 2,3$, the prediction accuracy for the $73$ bus system is comparable to that of the others. The low performance at $K = 1$ may be due to higher number of active sets with high frequency of occurrence as mentioned previously.

\section{Conclusions and Future Work}  \label{sec:conclusion}
This manuscript presents a neural-network (NN) based classifier for use in a data-driven optimal power flow solver. Instead of predicting the OPF solution directly from data, our NN classifier first determines the corresponding active set of constraints given an input set of uncertain injections. These active sets can then be used to determine the OPF solution using simple matrix inversion. The highlight of our methodology is the computational efficiency of model training and testing which make it amenable for real-time applications like corrective control. Using simulations with multiple test-cases and training set sizes, we demonstrate the efficacy of our approach. In particular, we show that a multi-ranked classifier (with top $3$ ranks) is able to achieve $99\%$ accuracy in active set prediction and corresponding OPF estimation. \\ 

Our active set classification approach is well-suited for extension in areas such as grid resilience, emergency operations as well as stochastic optimization, where identifying critical constraints is of paramount importance both from the perspective of operational security and solution efficiency. We are currently analyzing performance trade-offs for more complicated OPF problems including those with non-linear AC and binary constraints, where active sets can be more numerous than in DC-OPF. While we focus on active set classification, one can also build a set of parallel binary classifiers to predict the status of each individual constraint separately. It will be of interest to use our classification approach to develop a deeper understanding of various operational patterns in a given system, such as clustering of constraints that are typically simultaneously active/inactive, to better assist in development of tools for operational planning and real-time control.
\bibliographystyle{IEEEtran}
\bibliography{pscc2018.bib,references.bib}



\end{document}